\documentclass[twoside,12pt]{article} 
\begin{document}
\title{Low energy beta-beams}

 \author{C. Volpe\footnote{Proceedings to the 'European Strategy for
Future Neutrino Physics', 1-3 October 2009, CERN},$^{1}$
\\
$^1$Institut de Physique Nucl\'eaire, F-91406 Orsay cedex, \\
CNRS/IN2P3 and University of Paris-XI, France}

\maketitle

\begin{abstract}
The main goal of a beta-beam facility 
is to determine the possible existence 
of CP violation in the lepton sector, the value of the third
neutrino mixing angle and the mass hierarchy. 
Here we argue that a much broader physics case can be covered since
the beta-beam concept can also be
used to establish a low energy beta-beam facility. We discuss that
the availability of
neutrino beams in the 100 MeV energy range offers a unique
opportunity to perform neutrino scattering experiments of interest for
nuclear physics, for the study of fundamental interactions and of
core-collapse supernova physics.
\end{abstract}
 
\section{Introduction}
Beta-beams is a new concept for the production of neutrino beams that
is based on the beta-decay of boosted radioactive ions, as first
proposed by Zucchelli \cite{Zucchelli:2002sa}. By exploiting the high ion intensities
foreseen in the future, this method can produce intense neutrino beams, pure in
flavour and with well known fluxes. 
Such a method can also be used to establish a low energy beta-beam facility, as first proposed in \cite{Volpe:2003fi}.
Having neutrino beams in the 100 MeV energy range offers a unique
opportunity to perform experiments of interest for different domains
of physics, from nuclear physics to high energy physics and neutrino
astrophysics. (For a review on the beta-beam scenarios proposed so far, see \cite{Volpe:2006in}).

An important application of low energy beta-beams is neutrino-nucleus scattering 
measurements \cite{Volpe:2003fi,Serreau:2004kx,McLaughlin:2004va,Lazauskas:2007bs}. 
The available experimental data 
only include measurements on three nuclei, namely carbon, deuteron and iron. 
The results on carbon comprise both exclusive (involving the ground state to ground state transition)
and inclusive (involving ground state to excited states transitions) cross sections with decay-at-rest muons.
Even in this case, which the best studied one, the inclusive cross sections are not yet fully
understood \cite{Volpe:2000zn}. 
Deuteron represent a special case since predictions based on Effective Field Theories
reach few percent precision \cite{Nakamura:2002jg}. For iron, one  experiment has been performed with a low
precision, so that cross sections that differ up to a factor of 2 at a given energy
can fit the data \cite{Samana:2008pt}.  For all other nuclei, one has to rely upon theoretical predictions
that present significant differences, depending on the details of the model used. 

At present a precise knowledge of the nuclear response to neutrinos is essential
for a variety of timely applications, among which the search for
neutrinoless double-beta decay \cite{Ejiri,Volpe:2005iy} and the detection of (relic) supernova
neutrinos. Neutrino-less double-beta decay is the
only practical way to know about the neutrino nature (Majorana versus
Dirac). Experimentally we enter a phase in which the claim for its
evidence will be confirmed/refuted and the sensitivity on the
effective neutrino mass will be significantly lowered.
On the theoretical side, attempts are made to pin down the origin of the
theoretical discrepancies among the available calculations on the neutrino-less double-beta decay half-lives \cite{Rodin:2007fz}. However
much progress still needs to be made (see also \cite{Barea:2009zz}). One way to constrain the
half-life predictions is by using related processes,
namely beta-decay, muon capture, charge-exchange reactions and the (2$\nu$) double-beta decay.  

It has been discussed in the literature that neutrino-nucleus interaction measurements
can also furnish an indirect constrain to the half-life predictions \cite{Ejiri,Volpe:2005iy}. Indeed one can
show that the two-body matrix elements associated with the exchange of a Majorana neutrino
can be rewritten as product of the one-body matrix elements involving the same operators as those in neutrino-nucleus scattering \cite{Volpe:2005iy}. Therefore systematic neutrino-nucleus cross
section measurements can bring an important progress on our present knowledge of the involved nuclear response. Although it is clear that even with the possible future
high intensity neutrino beams, measurements on the candidate emitter can not be performed,
neutrino-nucleus scattering experiments can help in shading light on important questions that remain open. A key example is the generic feature, observed in all nuclei, that the measured Gamow-Teller matrix elements are
always smaller than the predictions. This is known in the literature as the
"quenching" problem and is usually taken into account by employing
an effective axial-vector coupling constant. Such a problem is particularly 
important for double-beta decay because it is a second-order process. 
While the origin of the quenching keeps since (many decades) not fully understood, it might also affect transitions of higher multipolarity that give a significant contribution to the neutrinoless double-beta decay half-lives. In this respect low energy beta-beam are an attractive tool because the neutrino fluxes can be varied by modifying the ion Lorentz boosts. Since the contributions 
from the states of different multipolarity vary with the average neutrino energy, one could disentangle them by combining measurements with different $\gamma$  \cite{Volpe:2003fi}.
A detailed analysis of the states contributing to the neutrino-nucleus cross sections with low energy beta-beams is made for lead in 
\cite{McLaughlin:2004va} and for several nuclei in \cite{Lazauskas:2007bs}, in comparison with the case of conventional sources. 

Gaining a precise description of the nuclear response to neutrinos is also a key step
for a precise knowledge of the neutrino detectors response when they use nuclei as targets,
e.g. oxygen, carbon,
argon, iron and lead. These are (or will be) used in particular to observe neutrinos
from core-collapse supernovae or to detect the diffuse supernova neutrino background
(for recent review on core-collapse supernova neutrinos and their relics see \cite{Duan:2009cd} and \cite{Ando:2004hc} respectively).  
An example is given by the HALO detector that is now planned at SNOLab. This exploits
neutrino-lead scattering in coincidence with one or two neutron emission. Indeed it
has been shown that this has an interesting sensitivity to the value of the third 
neutrino mixing value \cite{Engel:2002hg}. For this goal, a very precise measurement of the 
neutrino-lead scattering cross section is required. On the other hand the (main) detection
channel of some of the large-size detectors at present under study is through neutrino-nucleus scattering, as e.g. in the case of liquid argon detectors \cite{Autiero:2007zj}. In \cite{Lazauskas:2009yh}
it is pointed out that the search for relic galactic neutrinos, through the measurement of the Technetium abundance in Molybdenum ore, require a precise determination of the relevant cross sections. Another application of low energy beta-beams for core-collapse supernova physics is discussed in \cite{Jachowicz:2006xx}. Indeed it has been shown that by combining low energy neutrino scattering measurements with different ion boosts one could
fit the neutrino flux from a supernova explosion. Such a method could in principle be used to 
extract information on neutrino oscillations without the uncertainties from the cross sections \cite{Jachowicz:2006xx,Jachowicz:2008kx}. 

The study of fundamental interactions is another interesting research axis at a low energy beta-beam facility. Several cases have been discussed in the literature, including
a measurement of the Weinberg angle at low momentum transfer \cite{Balantekin:2005md} and a new test of the Conserved Vector Current (CVC) hypothesis \cite{Balantekin:2006ga}. The neutrino magnetic moment could also be investigated, although that requires a different configuration \cite{McLaughlin:2003yg}. As far as the Weinberg angle is concerned, the experiment exploits neutrino-electron scattering. The use of different Lorentz boosts can help reaching $10 \%$ precision if the systematic errors can be kept as low as 
10 $\%$ \cite{Balantekin:2005md}. This would improve the present precision by about 
a factor of 2. The CVC test is based on electron anti-neutrino scattering on protons in a Cherenkov (or scintillator) detector. In particular, it has been pointed out that the angular 
distribution of the emitted positrons is particularly sensitive to the weak magnetism contribution. So far this term has only been tested in beta-decay experiments in mirror nuclei. The calculations performed indicate that, after one year measurement, the weak magnetism term can be determined with a precisions of 9 $\%$ (1$\sigma$) if the systematic errors are kept as low as 5 $\%$ \cite{Balantekin:2006ga}. The sensitivity to physics beyond the standard model is discussed in \cite{Barranco:2007tz}. Finally, the authors of 
\cite{Bueno:2006yq} have analyzed the possibility to measure coherent neutrino-nucleus elastic scattering.

A key issue concerns the feasibility of low energy beta-beams.Two appear to be the ways to realize low energy beta-beams. One way is by storing the ions boosted at $\gamma=5-15$ in a devoted storage ring, as first proposed in \cite{Serreau:2004kx}. A preliminary study of such a storage ring has been addressed within the EURISOL Design Study, just ended (FP6) \cite{Payet}.  The second possibility is to put one/two detectors at off-axis of the storage ring planned for the CP violation search. Indeed the low energy component of the neutrino fluxes at $\gamma=100$ can be extracted by a substraction method, as pointed out in \cite{Lazauskas:2007va}. An off-axis configuration is discussed in \cite{Amanik:2007zy} as well.

In conclusion, there is a rich physics program that can be realized if a low energy neutrino facility in the 100 MeV energy range become available in the future, either with conventional sources or with beta-beams. Note that, if intense conventional neutrino sources become available at CERN, part of the physics program discussed here can be realized as well. In this respect low energy beta-beams have the attractive asset that the neutrino fluxes can be varied by changing the ion boost. This specific feature significantly increases the physics reach.

\end{document}